\documentclass[11pt, fleqn]{article}
\def\[{\begin{equation}}
\def\]{\end{equation}}

\usepackage{amsfonts, pstcol,times,color,graphicx,graphics,pstricks,pst-node,pst-coil,pst-grad, pst-plot, pst-3d, pst-fill, multido, pst-poly, url,bm, pst-math, pst-xkey, pstricks-add, pst-eucl, hyperref, pst-func}
\newpsobject{showgrid}{psgrid}{subgriddiv=1,griddots=10,gridlabels=6pt}

\begin{document}

\title{An exercise on Gauss' law for gravitation: The Flat Earth model }
\author{ A C Tort\footnote{email: tort@if.ufrj.br.}\\
Instituto de F\'{\i}sica
\\
Universidade Federal do Rio de Janeiro\\
Caixa Postal 68.528; CEP 21941-972 Rio de Janeiro, Brazil}
\maketitle
\begin{abstract}
We discuss the flat and hollow models of the Earth as a pedagogical example of the application of Gauss' law to the gravitational field. 
\end{abstract}
\vfill
\pagestyle{myheadings}
\markright{ \hskip 15cm  a c tort  2013}

\newpage

Few people still believe that the Earth is flat or is hollow. Or flat \emph{and} hollow.  
Among the models available some flat-earthers subscribe to the notion that the Earth is a disc whose center is at the North Pole. On the boundary of this disc there is a thick wall of ice, the Antartica,  that prevent the waters of going over the border. Notice that there is no South Pole in this model. There are other variants of this flat Earth model, e.g., the terrestrial disc could lie on an infinite plane. The Flat Earth model  is a good exercise on the application of Gauss' law  in a gravitational context. 
\vskip 10pt
Suppose we want to know how the gravitational field varies inside and outside the flat Earth. Gauss' law is the easiest way the answer these questions. Gauss' law for gravitation reads
\[
\Phi_{g}=-4\pi \, G M ,
\]
where $\Phi_{g}$  is the flux of the gravitational field through a closed smooth surface  $S$,  $G$  is the Newtonian gravitational constanat and  $M (S)$  is the mass enclosed by  $S$. If the mass distribution has a high degree of symmetry, e.g., spherical, cylindrical or planar then the flux can be easily calculated provided that we choose a Gaussian surface that respects the symmetry of the configuration at hand. If this is the case the gravitational flux  is  given by

\[\Phi_{g}= - g A ,\]
where $g$ is the magnitude of the field on the Gaussian surface whose area is $A$. The minus sign is due to the fact that the Gaussian surface has an orientation. The unit normal vector on any point on this surface points outwards and because $\mathbf{g}$ always points inwards it follows that the flux is also always negative. 

\begin{figure}[!h]
\begin{center}
\begin{pspicture}(-8,-3)(8,6)
\psset{arrowsize=0.2 1}
\pspolygon[linestyle=none, fillstyle=solid, fillcolor=lightgray](-6,0)(-6, 3 )(6,3)(6,0)
\psline[linewidth=0.25mm, linestyle=dashed](-6,1.5)(6, 1.5)
\pcline[linewidth=0.15mm, offset=-15pt]{|-|}(6,0)(6,3)
\lput*{:L}{$H$}
\pspolygon[linewidth=0.35mm, linestyle=dashed](2,-1)(2, 4 )(4,4)(4,-1)
\pspolygon[linewidth=0.35mm,linestyle=dashed](-4,1)(-4, 2 )(-1, 2)(-1,1)
\psline[linewidth=1.25pt, linestyle=solid, linecolor=red]{->}(-2.5, 2.15)(-2.5, 3.00)
\psline[linewidth=1.25pt, linestyle=solid, linecolor=blue]{->}(-2, 4)(-2,  2)
\psline[linewidth=1.25pt, linestyle=solid, linecolor=red]{->}(-2.5, 0.85)(-2.5, 0.0)
\psline[linewidth=1.25pt, linestyle=solid, linecolor=blue]{->}(-2, -1)(-2,  1)
\psline[linewidth=1.25pt, linestyle=solid, linecolor=red]{->}(3, 4.15)(3, 5.00)
\psline[linewidth=1.25pt, linestyle=solid, linecolor=blue]{->}(3.5, -3)(3.5,  -1)
\psline[linewidth=1.25pt, linestyle=solid, linecolor=red]{->}(3, -1.15)(3, -2)
\psline[linewidth=1.25pt, linestyle=solid, linecolor=blue]{->}(3.5, 6)(3.5,  4)
\rput(-3,2.5){$\mathbf{\hat n}$}
\rput(-3.25,0.5){$- \mathbf{\hat n}$}

\rput(2.5,4.5){$\mathbf{\hat n}$}
\rput(2.25,-1.5){$- \mathbf{\hat n}$}

\rput(4,-2){$ \mathbf{g}$}
\rput(4,5){$- \mathbf{g}$}

\rput(-1.5,3){$- \mathbf{g}$}
\rput(-1.5, 0){$ \mathbf{g}$}

\rput(-6.85, 3){$x=\frac{H}{2}$}
\rput(-6.85, 1.5){$x=0$}
\rput(-6.85, 0){$x=-\frac{H}{2}$}

\rput(0.5, 2){$\rho$}
\rput(-4, 3.35){$\sigma$}

\rput(-4.5, 1){$S_1$}
\rput(1.5, 1){$S_2$}
\end{pspicture}
\caption{Flat Earth and Gaussian surfaces.}
\label{FEGS}
\end{center}
\end{figure}
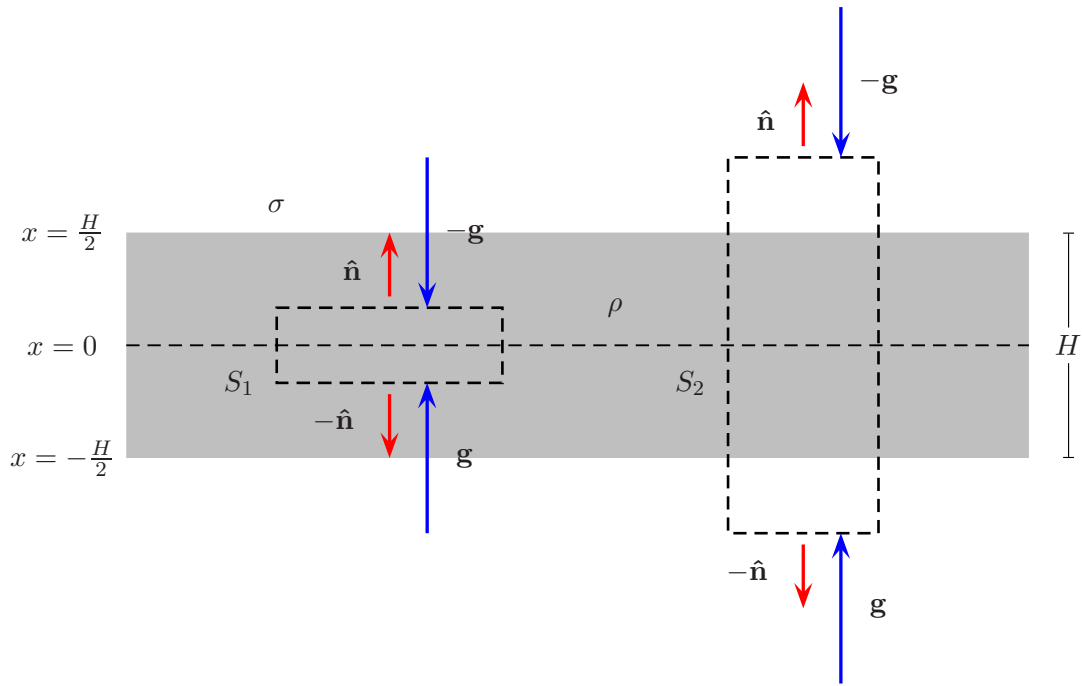

\vskip 10pt
Though flat-earthers do not believe in gravity we round-earthers do hence let us see how Gauss' law applies to the flat Earth model\footnote{The problem we are about to discuss can be approached by considering its electrostatic counterpart as in \cite{Riley2001}. Here in order to emphasize the notion that Gauss' law holds ffor gravitation, in fact for any field that depends on $1/r^2$ we chose to start from its formulation for the gravitational field from the beginning.}. First of all we must realize that the terrestrial disc is really a highly flattened cylinder. This means that if $a$ is the radius of the cylinder and $H$ is its height, or better, its thickness, then the condition $a \gg H$ holds. If we additionally agree to consider points far away from the border of the cylinder then planar symmetry applies provided that the mass distribution $\rho$ is uniform or a function of the thickness of the (flat) Earth only. Notice that this means that the field is perpendicular to the mass distribution. For simplicity will suppose also that $\rho$ is uniform and its numerical value equal to the mean density of the spherical Earth. If we adopt these assumptions we can adopt the cylindrical surfaces $S_1$ and $S_2$ as Gaussian surfaces as sketched in Figure \ref{FEGS}.  Consider for example $S_1$.  The flux through this surface is 

\[ \Phi_{g}= - 2 g A_{\mbox{\tiny top}} ,\]
and the mass enclosed by $S_1$ is
\[M(S_1)  = \rho A_{\mbox{\tiny top}} x .\] 
Gauss' law then leads to

\[ g(x) =  4\pi\, G\rho \, x,   \]
for the field inside the distribution.  For the evaluation of the field outside the distribution we make use of $S_2$. Then, proceeding in the same way we find $\Phi_g = -2 g  A_{\mbox{\tiny top}}$, but this time the mass enclosed is $M(S_2)=\rho H A_{\mbox{\tiny top}}$.  It follows from Gauss' law that   in magnitude outside the mass distribution $g=2\pi \,G\sigma$, where we have deifined  $\sigma=\rho H$  as mean surface density of the (flat) Earth. We can collect these results taking into account their domain of validity and direction in the formula given below 
(notice that the field is continuous on the surface of the distribution)
\[ g(x) = - \left\{ \begin{array}{c} 2\pi \,G\sigma , \hskip 10 pt x \leq - \frac{H}{2};\\ \\ - 4\pi\, G\rho \, x, \hskip 10pt -\frac{H}{2}\leq x + \leq \frac{H}{2}; \\ \\ - 2\pi\, G\sigma , \hskip 10 pt x \geq + \frac{H}{2}\end{array} \right. ,\]

\vskip 10pt
And what if the Earth is flat and hollow? In this case the reader can easily verify that gravitational field reads

\[ g(x) = \left\{ \begin{array}{c} 4\pi \,G\sigma , \hskip 10 pt x \leq -\frac{H}{2}; \\ \\ 0 , \hskip 10pt -\frac{H}{2}\leq x \leq \frac{H}{2}; \\ \\ - 4\pi\, G\sigma , \hskip 10 pt x \geq \frac{H}{2}\end{array} \right. .\]
\vskip 10pt
The Flat Earth Model or the Hollow Flat Earth Model must reproduce the measured value of the gravitational acceleration  on the surface of the Earth.  In the case of the former model  it is reasonable to ask ourselves how its  thickness $H$ compares to the radius $R$ of the spherical Earth \cite{Riley2001}.  The mass enclosed by the Gaussian cylinder will be  $ M = \rho \, A H\,$ where as mentioned before $\rho$ is the mean density of the spherical Earth.  On the surface of the (flat) Earth

\[ g = 2\pi G \, \rho H . \label{gdisc} \]
Since $g$ must be equal to the gravitational acceleration on the surface of the (spherical) Earth we have

\[ g=  \frac{G \, M_{\mbox{\tiny Earth }}}{R^2}=G\,\frac{4\pi R \rho }{3} ,\label{gsphere} \]
where $R$ is the mean radius of the terrestrial sphere. Setting  Eq.  (\ref{gdisc}) equal to Eq. (\ref{gsphere}) it follows that

\[ H= \frac{2}{3}\,R  = \frac{2}{3}\, 6 371 \hskip 2.5pt \mbox{km} \approx 4 \, 247 \hskip 2.5pt \mbox{km}. \label{H}\]
Notice that this result does not depend on the mean density of the Earth. If the Earth were a \lq lighter\rq  or \lq heavier\rq \hskip 5pt planet this result would still hold. As a consistency check the reader can infer the value of $\rho$ from the measured value of $g$ on the surface of the Earth. 
\vskip 10pt
The thickness of the flat Earth can be also inferred from experimental data. Suppose we measure $g$ on the surface of the Earth and find $9. 807\,$m/s$^2$, and the from rock samples we conclude that the mean mass density $\rho$ is $5\,515\,$ kg/m$^3$. Then from Eq. (\ref{gdisc}) we can write

\[H = \frac{g}{2\pi G \rho}= \frac{9. 807}{2\pi \, 6.673 \,\times\, 10^{-11} \,\times \, 5. 515  }\,\mbox{m} = 4\, 241\, 208. 91\,\mbox{m}  \approx 4241\,\mbox{km} ,\]
in good agreement with the theoretical result given by Eq. (\ref{H}). Suppose flat Earth engineers decide to bore a tunnel to communicate with flat-earthers on the other side of the world. A tunnel boring machine working round the clock  can progress more or less $15$ meters/day. A simple calculation will show that the engineers  will need $775$ years to reach the other side of the Earth!
\vskip 10pt
Flat Earth models have a long history and there are many of them. Here we have discussed just  two of those models. The reader can find more information on flat  Earth theories at, for example
\vskip 10pt
 \hskip 5pt \href{http://www.lhup.edu/~dsimanek/flat/flateart.htm}{\textbf{\small \blue http://www.lhup.edu/~dsimanek/flat/flateart.htm}}.  
 \vskip 10pt
 See also Sir Patrick Moore's \emph{Can You Speak Venusian?} for a delightful point of view on this subject \cite{Moore1972}.  If the reader wants the real thing she/he can try for example
 
 \vskip 10pt
 \hskip 5pt \href{http://theflatearthsociety.org/cms/}{\textbf{\small \blue http://theflatearthsociety.org/cms/}}.  
 \vskip 10pt


\end{document}